\documentclass[twocolumn, 10pt]{IEEEtran}

\usepackage{hyperref,bm,graphicx,amsmath,amssymb}

\begin{document}
\title{Bayesian Hypothesis Test for sparse support recovery using Belief Propagation}

\author{ Jaewook~Kang,
        Heung-No~Lee,
        and~Kiseon~Kim\\
\bigskip
\IEEEauthorblockA{School of Information and Communication,\\
 Department of Nanobio Materials and Electronics,\\
 Gwangju Institute of Science and Technology (GIST), Gwangju 500-712, South Korea\\
(Tel.: +82-62-715-2264, Fax.:+82-62-715-2274,
Email:\{jwkkang,heungno,kskim\}@gist.ac.kr})

\thanks{This work was proceeded in \emph{IEEE Statistical Signal Processing Workshop} (SSP), Ann Arbor, Aug. 2012}

 }\maketitle

\begin{abstract}
In this paper, we introduce a new support recovery algorithm from
noisy measurements called \emph{Bayesian hypothesis test via belief
propagation} (BHT-BP). BHT-BP focuses on sparse support recovery
rather than sparse signal estimation. The key idea behind BHT-BP is
to detect the support set of a sparse vector using hypothesis test
where the posterior densities used in the test are obtained by aid
of belief propagation (BP). Since BP provides precise posterior
information using the noise statistic, BHT-BP can recover the
support with robustness against the measurement noise. In addition,
BHT-BP has low computational cost compared to the other algorithms
by the use of BP. We show the support recovery performance of BHT-BP
on the parameters $(N,M,K,\text{SNR})$ and compare the performance
of BHT-BP to OMP and Lasso via numerical results.
\end{abstract}

\section{Introduction}\label{sec:intro}
Support recovery (also known as sparsity pattern recovery) refers to
the problem of finding the support set corresponding to $K$ nonzero
elements of a sparse vector  $\mathbf{x} \in \mathbb{R}^N$ from a
noisy measurement vector $\mathbf{z} \in \mathbb{R}^M$. This problem
is fundamentally important to solve underdetermined systems $(M <
N)$ because once the support set is known, the system is simplified
to an overdetermined system, which can be solved by the conventional
least square approach. Therefore, the support recovery is associated
with a broad variety of underdetermined problems such as compressed
sensing \cite{CS}, subset selection in regression \cite{subset_sel},
sparse approximation \cite{sparse_app}.

Recently, a plenty body of works has analyzed the performance of the
support recovery \cite{Wainwright1}-\cite{Akcakaya}. Wainwright
investigated the performance of  the support recovery with Lasso in
\cite{Wainwright1}, and Fletcher and Rangan analyzed that of OMP in
\cite{Fletcher1}. In addition, the theoretical limit of the support
recovery has been discussed  in terms of \emph{maximum likelihood}
(ML) \cite{Fletcher2},\cite{Rad}, and information theory
\cite{Wainwright2},\cite{Akcakaya}.  These theoretical works reveal
that current practical algorithms have a potentially large gap from
the theoretic limit; therefore, developing practical algorithms
which approach the limit is an open challenge.

One line of approaches is sparse recovery using sparse matrices
\cite{CS-BP}-\cite{CS-BSD}. These works are inspired by the success
of \emph{low-density parity-check} codes
\cite{Gallager},\cite{Richardson}.  The use of the sparse matrix
enables simple and fast measurement generation. In addition, these
approaches can be made more attractive if they are applied in
conjunction with belief propagation (BP). BP replaces the recovery
process by iterative message-passing processes. This replacement
reduces the computational cost to the $O(N \log N)$ order
\cite{CS-BP}.

Motivated by such previous works, in this paper, we propose a new
support recovery algorithm using BP called \emph{Bayesian hypothesis
test via BP} (BHT-BP). BHT-BP utilizes a hypothesis test to detects
the support set from noisy measurements, where the posterior density
used in the test is provided by BP. Hence, BHT-BP has low
computational cost from the use of BP and noise robustness from the
hypothesis test.

 In our previous work \cite{CS-BSD}, BHT-BP was a part to
provide support set information for the process of sparse signal
estimation. Differently from \cite{CS-BSD}, this paper aims to
investigate performance of support recovery by BHT-BP rather than
the signal estimation. We show the support recovery performance of
BHT-BP on the parameters $(N,M,K,\text{SNR})$ and demonstrate the
superiority of BHT-BP compared to support recovery by OMP and Lasso
via simulation results.

\section{Problem Formulation}
\subsection{Signal Model}
Let $\mathbf{x} \in \mathbb{R}^N$ denote a random $K$-sparse vector
with a state vector $\mathbf{s(x)}$ indicating the support set of
$\mathbf{x}$; therefore $||\mathbf{s(x)}||_0=K$. Each element $s_i
\in \mathbf{s(x)}$ is defined as
\begin{eqnarray}\label{eq:eq2-1}
{s_i} = \left\{ \begin{array}{l}
1,\,\,\,\,\,{\rm{if}}\,\,{x_i} \ne 0\\
0,\,\,\,\,\,{\rm{else}}
\end{array} \right.\text{ for all }\,\, i \in \{1,...,N\}.
\end{eqnarray}
We limit our discussion to the random vector $\mathbf{x}$ whose
elements are i.i.d. random variables. Then, the decoder observes a
measurement vector $\mathbf{z} \in \mathbb{R}^M$, given as
\begin{eqnarray}\label{eq:eq2-2}
\mathbf{z}=\mathbf{\Phi}\mathbf{x}_0 + \mathbf{n},
\end{eqnarray}
where $\mathbf{x}_0 \in \mathbb{R}^N $ is a deterministic
realization of $\mathbf{x}$; and $\mathbf{n} \sim
\mathcal{N}(0,\sigma_n^2\mathbf{I}_M )$ is an additive Gaussian
noise vector. For the measurement matrix $\mathbf{\Phi}$, we employ
sparse-Bernoulli matrices $\mathbf{\Phi}\in\{0,1,-1\}^{M \times N}$
with rank($\mathbf{\Phi}) \leq M$ and $M \leq N$. Namely, in the
matrix, sparsely nonzero elements are equiprobably equal to $1$ or
$-1$. In addition, we fix the column weight of $\mathbf{\Phi}$  to
$L$ such that $\left\|{\mathbf{\phi}_{jth\,\, col}}\right\|_2^2=L$.

\subsection{Problem Statement}
The goal of the decoder is to detect the support set
$\mathbf{s}(\widehat{\mathbf{x}}_0)$ from $\mathbf{z}$ where each
supportive state $s_i(\widehat{x}_{0,i})$ is independently detected
in each element unit, given as:
\begin{eqnarray}\label{eq:eq2-3}
{\frac{{\Pr \{ s_i  = 0|{\bf{z}}\} }}{{\Pr \{ s_i  = 1|{\bf{z}}\}
}}} \mathop {\mathop \gtrless \limits_{{H_1}} }\limits^{{H_0}} 1
\text{ for all } i \in \{1,...,N\},
\end{eqnarray}
where $H_0:s_i(x_{0,i})=0$ and $H_1:s_i(x_{0,i})=1$ are two possible
hypotheses.

To see the performance of the algorithms, we measure the state error
rate (SER) between the detected support
$\mathbf{s}(\widehat{\mathbf{x}}_0)$ and the true support
$\mathbf{s}(\mathbf{x}_0)$, given as
\begin{eqnarray}\label{eq:eq2-4}
{\text{SER}}: = \frac{{\# \{ i \in
\{1,...,N\}|s_i(\widehat{x}_{0,i}) \ne s_i(x_{0,i}) \} }}{N}.
\end{eqnarray}
We are interested in the SER performance as a function of
undersampling ratio $M/N$ for a veriety of signal sparsity $K$ and
SNR defined as
\begin{eqnarray}\label{eq:eq2-5}
\text{SNR :} =10 \log_{10} \frac{{E{\left\| {{\bf{\Phi}} \mathbf{x}}
\right\|_2^2 } }}{{M\sigma _{n }^2 }} \text{ dB }.
\end{eqnarray}

\section{Proposed Algorithm}

\subsection{Prior Model}
 We first specify our prior model  since the proposed algorithm is
derived on the basis of the Bayesian rule.  By associating state
variable $s_i$, we model the prior density of $x_i$ using a
\emph{spike-and-slab} model originating in a two-state mixture
density as follows:
\begin{eqnarray} \label{eq:eq3-1}
f_{x}(x)&:=& qf_{x}( x |s = 1)+ (1 -
q)f_{x}( x |s = 0)\nonumber\\
 &=& q\mathcal{N}(x;0,\sigma _x^2 )+ (1 - q)\delta (x ),
\end{eqnarray}
where $\delta(x)$ indicates a Dirac distribution having nonzero
value between $x \in [0-, 0+]$ and $\int \delta(x)dx =1$;
$q=\frac{K}{N}$ is the sparsity rate. In addition, we drop the index
$i$ from the prior density under the assumption of i.i.d. elements.
The spike-and-slab prior has been widely employed in Bayesian
inference problems \cite{Ishwaran}.

\subsection{Hypothesis Detection of Support}
\begin{figure}[!t]
\centering
\includegraphics[width=9cm]{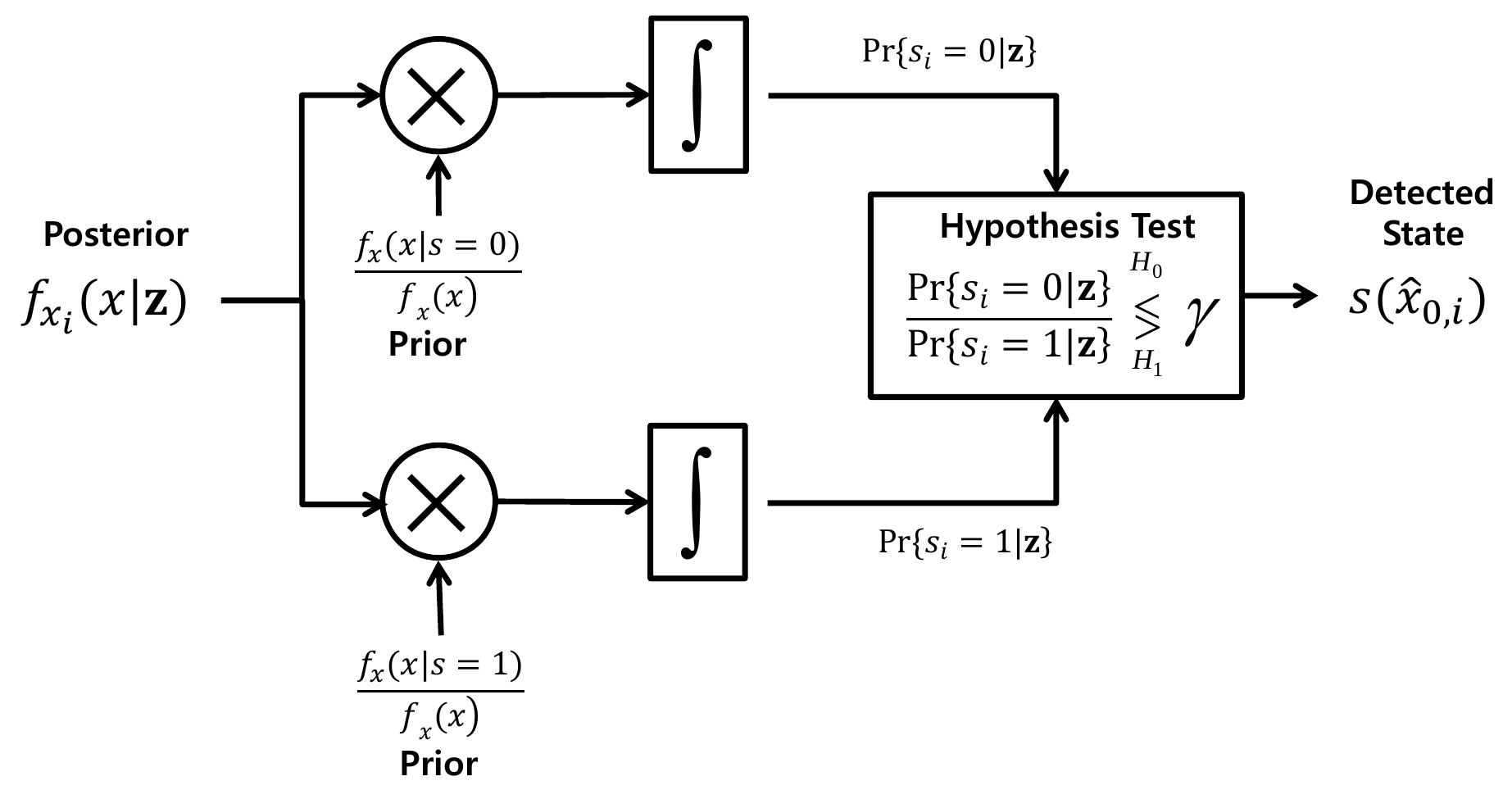}
\caption{{\small{Bayesian hypothesis test for a state detection}}}
\label{fig:Fig3-1}
\end{figure}

In order to perform the hypothesis test in \eqref{eq:eq2-3}, the
decoder needs to calculate a probability ratio $ \frac{{\Pr \{ s_i =
0|{\bf{z}}\} }}{{\Pr \{ s_i = 1|{\bf{z}}\} }} $. By factorizing over
$x_i$, the ratio becomes
\begin{eqnarray}\label{eq:eq3-2}
{\frac{{\Pr \{ s_i  = 0|{\bf{z}}\} }}{{\Pr \{ s_i  = 1|{\bf{z}}\}
}}}=  {\frac{{\int  {\Pr \{ s_i  = 0|{\bf{z}},x_i \} f_{x_i}(x
|\mathbf{z} ) }dx }}{{\int  {\Pr \{ s_i = 1|{\bf{z}},x_i \} f_{x_i}(
x |\mathbf{z} ) }dx }}} \mathop {\mathop \gtrless \limits_{{H_1}}
}\limits^{{H_0}} 1,
\end{eqnarray}
where $f_{x_i}( x |\mathbf{z} )$ denotes the posterior density of
$x_i$ given $\mathbf{z}$. In \eqref{eq:eq3-2}, $\Pr \{ s_i
|{\bf{z}},x_i \} = \Pr \{s_i |x_i \}$ holds true since the
measurements $\mathbf{z}$ do not provide any additional information
on $s_i$ given $x_i$. Using the Bayesian rule and the prior
information, the test in \eqref{eq:eq3-2} is rewritten as
\begin{eqnarray}\label{eq:eq3-3}
{\frac{{\int  {\frac{{f_{x}(x |s = 0) }}{{f_{x}(x ) }}f_{x_i}(x
|\mathbf{z} )}dx }}{{\int {\frac{{f_{x}( x |s = 1) }}{{f_{x}( x )
}}f_{x_i}(x |\mathbf{z} ) }dx }}} \mathop {\mathop \gtrless
\limits_{{H_1}} }\limits^{{H_0}} {\frac{{\Pr \{ s = 1\} }}{{\Pr \{ s
= 0\} }}} =  \gamma,
 \end{eqnarray}
where $\gamma:={\frac{q}{(1-q)}}$. Therefore, the probability ratio
is obtained from the corresponding posterior and prior densities.
The overall flow of the hypothesis test for a supportive state
detection is shown in Fig.\ref{fig:Fig3-1}.

\subsection{Belief Propagation for Posterior Approximation}
The posterior density $f_{x_i}(x|\mathbf{z})$ used in the hypothesis
test is obtained by BP. Using Bayesian rule, we can represent the
posterior density $f_{x_i}(x|\mathbf{z})$ in the form of $
{\rm{Posterior = Prior}} \times
\frac{{{\rm{Likelihood}}}}{{{\rm{Evidence}}}}$, given as
\begin{eqnarray}\label{eq:eq_12}
f_{x_i}(x|\mathbf{z}) =   f_{x}(x)\times \frac{ { f_{\mathbf{z}}(
{{\bf{z}}|x_i)}}}{f_{\mathbf{z}}(\mathbf{z})}.
\end{eqnarray}
If the measurement matrix $\mathbf{\Phi}$ is sufficiently sparse
such that the corresponding bipartite graph is tree-like, we
postulate that the elements of $\mathbf{z}$ associated with $x_i$
are independent each other \cite{Richardson}. Under the tree-like
assumption, we can decompose the likelihood density $f_{\mathbf{z}}(
{\bf{z}}|x_i)$ to the product of densities:
\begin{eqnarray} \label{eq:eq_13}
f_{x_i}(x|\mathbf{z}) \propto f_{x}(x) \times
\prod\limits_{j:\phi_{ji}\neq 0} f_{z_j}({z}|{x_i}).
\end{eqnarray}
Since each element of $\mathbf{z}$ is represented by the sum of
independent random variables with $\mathbf{\Phi}$, we can expressed
$f_{z_j}(z|x_i)$ as linear convolution of  associated density
functions, given as
\begin{eqnarray} \label{eq:eq3-4}
\begin{array}{l}
f_{z_j}(z|x_i)=\\
\delta(z - z_j) \otimes f_{n_j}(n) \otimes
\left(\bigotimes\limits_{k:\phi_{jk}\neq 0, k\neq i} {f_{x_k}}(x)
\right).
\end{array}
\end{eqnarray}

The essence of the BP-process is to obtain an approximation of
$f_{x_i}(x|\mathbf{z})$ from iterative mutual update of probability
messages. The message update rule is formulated based on
\eqref{eq:eq_13},\eqref{eq:eq3-4}. We follow the rule introduced in
\cite{CS-BSD}, given as
\begin{eqnarray}\label{eq:eq3-5}
\mathbf{a}_{i \rightarrow j}^l:=\eta\left[ {{f_{x}}(x)  \times
\prod\limits_{ k: \phi_{ki} \neq 0, k \neq j   } {{\bf{b}}_{k
\rightarrow i}^{l-1} } } \right],
\end{eqnarray}
\begin{eqnarray} \label{eq:eq3-6}
\mathbf{b}_{j \rightarrow i}^l:=  \left( {\mathop \bigotimes
\limits_{k:\phi_{jk}\neq 0, k\neq i}{\bf{a}}_{k \to j}^l} \right)
\otimes \delta(z - z_j) \otimes f_{n_j}(n) ,
\end{eqnarray}
$\text{ for all } (i,j) \in \{1,...,N\} \times \{1,...,M\}:
|\phi_{ji}|=1$, where $\otimes$ and $\bigotimes$ are the operator
for linear convolution and the linear convolution of a sequence of
functions, respectively; $\eta(\cdot)$ denote a normalization
function for probability densities;, and $l$ denotes the iteration
index. The probability messages $\mathbf{a}_{i \rightarrow j}$ and
$\mathbf{b}_{j \rightarrow i}$ are mutually updated via
BP-iterations. Then, we approximate the posterior density
$f_{x_i}(x|\mathbf{z})$ after a certain number of iterations $l^*$
as follows:
\begin{eqnarray} \label{eq:eq3-7}
f_{x_i}(x |\mathbf{z} ) \approx \eta\left[ { f_{x}(x) \times
\prod\limits_{j: \phi_{ji}\neq 0} {{\mathbf{b}_{j \rightarrow
i}^{l^*}} } } \right].
\end{eqnarray}

\section{Numerical Results} \label{Numresult}
We demonstrate the performance of BHT-BP via simulation results. We
consider the SER performance given in \eqref{eq:eq2-4} and take 1000
Monte Carlo trials for each experimental point to show the average
performance. In each trial, we generate a deterministic sparse
vector $\mathbf{x}_0$ with $N=128$. The generation of $x_i$
belonging to the support set follows zero mean Gaussian distribution
with $\sigma_x=10$; but we restrict the magnitude level to
$\sigma_x/5 \leq |x_i| \leq 3\sigma_x$. For the measurement matrix
$\mathbf{\Phi}$, BHT-BP uses sparse-Bernoulli matrix with $L=3$.

\begin{figure}[!t]
\centering
\includegraphics[width=9cm]{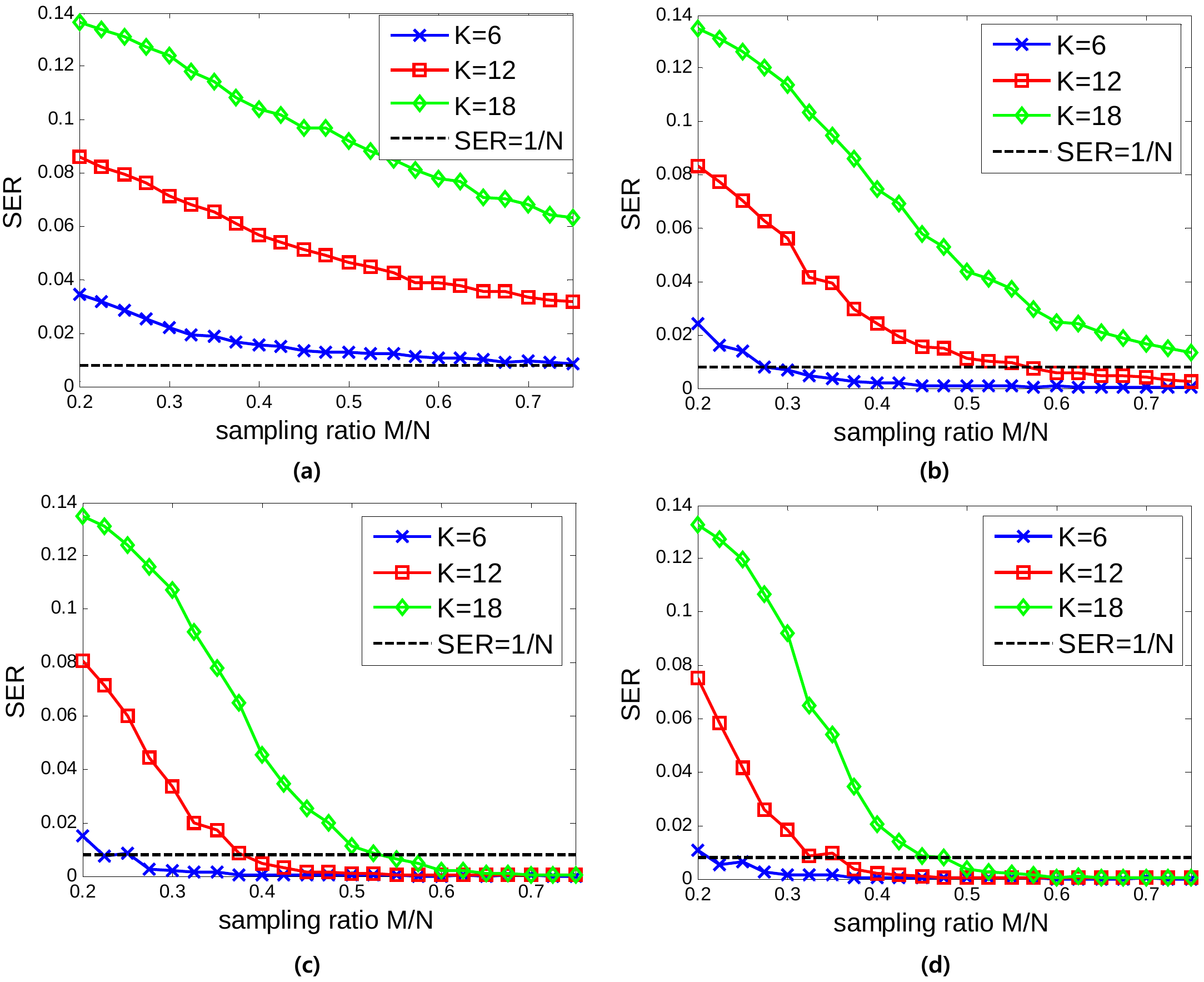}
\caption{{\small{SER performance of BHT-BP over $M/N$ for a variety
of SNR and $K$: (a) SNR=10 dB, (b) SNR=20 dB, (c) SNR=30 dB, (d)
SNR=50 dB.}}} \label{fig:Fig4-1}
\end{figure}

Fig.\ref{fig:Fig4-1} shows the SER performance as a function of
undersampling ratio $M/N$ for a variety of SNR and $K$. Naturally,
BHT-BP is well performed as SNR increases. However, BHT-BP works
similarly beyond 30 dB as shown in Fig.\ref{fig:Fig4-1}-(c),(d).
Regarding the signal sparsity $K$, BHT-BP needs at least SNR $\geq
30$ dB to recover the support with SER below the $1/N \approx
0.0078$ for $K \leq 18$ if  $M/N$ is sufficient.

\begin{figure}
\centering
\includegraphics[width=9cm]{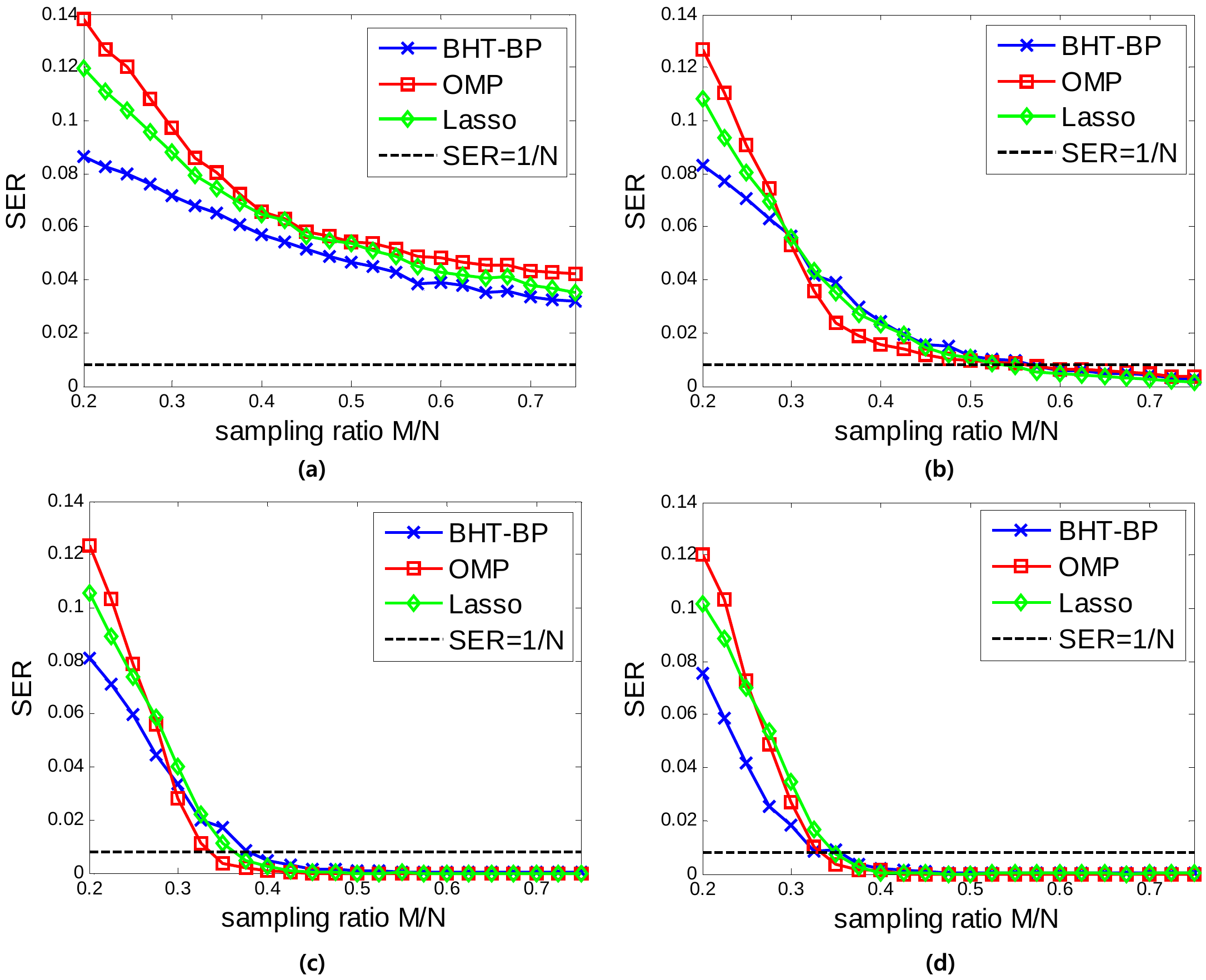}
\caption{{\small{SER performance comparison to OMP and Lasso over
$M/N$ when $K=12$: (a) SNR=10 dB, (b) SNR=20 dB, (c) SNR=30 dB, (d)
SNR=50 dB.}}} \label{fig:Fig4-2}
\end{figure}

Fig.\ref{fig:Fig4-2} shows the advantages of BHT-BP compared to OMP
\cite{OMP} and Lasso \cite{Lasso}. The source codes of OMP and Lasso
were obtained from \emph{SparseLab 2.0} package (available at
http://sparselab.stanford.edu/). OMP and Lasso use a Gaussian
measurement matrix having the same column energy as the
sparse-Bernoulli matrix for fairness, \emph{i.e.}, $\left\|
{\mathbf{\phi}
_{j,Gaussian}}\right\|_2^2=\left\|{\mathbf{\phi}_{j,Sparse}}\right\|_2^2=L$.
In addition, we choose the $K$-largest values from
$\widehat{\mathbf{x}}_0$ for the support detection of OMP and Lasso.
We generate $\mathbf{x}_0$ with $N=128$ and $K=12$ in this
simulation. At SNR=10 dB, we can see that BHT-BP outperforms Lasso
and OMP as shown in Fig.\ref{fig:Fig4-2}-(a), because the use of
noise statistic $f_{n_j}(n)$ in the BP-process provides the precise
posterior information, and it leads to reducing of misdetection of
the supportive state $\widehat{s}_{0,i}$ in the hypothesis test.
Nevertheless, the SER curves at SNR=10 dB are far from $SER=1/N$ due
to the noise effect. Such a fact is supported by the theoretical
work in \cite{Fletcher2} where the ML decoder needs $M/N \geq 1.02$
for exact support recovery when SNR=10 dB.

As SNR increases, all algorithms are gradually performed similarly,
but Lasso and OMP have slightly lower cross point to SER$=1/N$  as
shown in Fig.\ref{fig:Fig4-2}-(b),(c),(d) and Table \ref{table1}.
Such a result is caused by sensing inefficiency from the use of
sparse measurement matrices \cite{Wainwright2}. However, BHT-BP has
advantage on low computational cost and the fast measurement
generation with the sparse matrix. Indeed, BHT-BP has the lower cost
$O(N \log N)$ by aid of BP, than OMP $O(KNM)$ and Lasso $O(NM^2)$.

\begin{table}[!t]\label{table1}
\renewcommand{\arraystretch}{1.0}

\caption{Cross point in $M/N$ to SER$=1/N$ when $K=12$}
\label{table1}
 \centering

\begin{tabular}{||c||c|c|c|c||}
\hline\hline
Algorithms / SNR &  50 dB   & 30 dB     & 20 dB & 10 dB \\
\hline \hline
BHT-BP &     0.357    & 0.375     & 0.575 &   -  \\
\hline
Lasso  &     0.350    & 0.361     & 0.550 &   -  \\
\hline
OMP &        0.331    & 0.335     & 0.552 &  - \\
\hline
ML limit for exact  &  0.101 & 0.1108 & 0.1935 & 1.021 \\
recovery in \cite{Fletcher2}&    &      &       &\\
 \hline\hline
\end{tabular}
\end{table}

\section{Summary}
We proposed a new support recovery algorithm using belief
propagation called BHT-BP. Our proposed algorithm utilizes
hypothesis test to detects the support set from noisy measurements
where posterior used in the test is provided by belief propagation.
Our numerical results showed that the proposed algorithm outperforms
OMP and Lasso in the low SNR regime, and becomes working similarly
as SNR increases. However, the proposed algorithm still has strength
in terms of low computational cost and the fast measurement
generation by the use of the sparse matrix and belief propagation.

\section{Acknowledgment}
{\small{This work was supported by the National Research Foundation
of Korea (NRF) grant funded by the Korea government(MEST) (Haek-Sim
Research Program, NO. 2011-0027682, Do-Yak Research Program, NO.
2012-0005656)}}



\end{document}